\documentstyle[twocolumn,aps,epsfig]{revtex}

\begin{document}
\wideabs{ \title{Generation of Continuous Variable
Einstein-Podolsky-Rosen Entanglement\\
via the Kerr Nonlinearity in an Optical Fibre}

\author{Ch.~Silberhorn$^{1}$, P.~K.~Lam$^{1,2}$, O.~Wei{\ss}$^{1}$,
F.~K\"{o}nig$^{1}$, N.~Korolkova$^{1}$, and G.~Leuchs$^{1}$}

\address{$^{1}$Lehrstuhl f\"ur Optik, Physikalisches Institut der
Universit\"at Erlangen-N\"urnberg, \\
Staudtstra{\ss}e 7/B2, D-91058 Erlangen, Germany.\\
$^{2}$Department of Physics, Faculty of Science, The Australian
National University, ACT 0200, Australia.}

\date{\today}
\maketitle

\begin{abstract}
We report on the generation of a continuous variable
Einstein-Podolsky-Rosen (EPR) entanglement using an optical fibre
interferometer.  The Kerr nonlinearity in the fibre is exploited
for the generation of two independent squeezed beams.  These
interfere at a beam splitter and EPR entanglement is obtained
between the output beams.  The correlation of the amplitude
(phase) quadratures are measured to be $4.0 \pm 0.2 \ (4.0 \pm
0.4)$ dB below the quantum noise limit. The sum criterion for
these squeezing variances $0.80  \pm 0.03 < 2 $ verifies the
nonseparability of the state. The product of the inferred
uncertainties for one beam $(0.64 \pm 0.08)$ is well below the
EPR limit of unity.
\end{abstract}

}

\narrowtext

Since the original proposal of a {\it Gedankenexperiment}
intending to show the incompleteness of quantum mechanics in 1935
\cite{EPR}, a number of schemes for generating the
Einstein-Podolsky-Rosen (EPR) entanglement have been realized. The
schemes range from the production of gamma-ray pairs from
positron-electron annihilations \cite{Wu}, to proton pairs
\cite{Lamehi-Rachti}, to pairs of low-energy photons from atomic
radiative cascade \cite{Clauser} and more recently to schemes
involving optical parametric processes \cite{Shih}.  Most of
these initial experiments utilized the entanglement as originally
intended: to test the validity of quantum mechanics via either
the violation of Bell inequality \cite{Clauser} or the
demonstration of the EPR paradox \cite{Ou}.  Following the
proposals of a myriad of quantum information schemes in recent
years where entanglement is regarded as a basic requisite, the
subject matter has experienced a resurgence of interest. The
purposes of entanglement generation are now shifting to that of
quantum information applications. Amongst these applications are
the realization of quantum teleportation, the implementation of
dense coding, quantum cryptography  and other quantum
communication schemes \cite{world}.  In view of these changing
needs, it is desirable to explore simpler and more reliable
alternatives for the generation of EPR entanglement.

In this letter, we report on what is to our knowledge the first
generation of EPR entanglement of photons that does not rely on
any pair production process, such as those in the above-mentioned
examples. Instead, the Kerr $(\chi^{(3)})$ nonlinearity of an
optical fibre is utilized to produce two amplitude squeezed
beams, with the nonlinear interaction on each beam uncoupled to
the other.  To create the EPR entanglement, no additional
nonlinear interaction is required. Instead, the amplitude
squeezed beams are made to interfere at a 50/50 beam splitter
\cite{Loock}. In this vein sum squeezing is obtained for the
amplitude quadratures and difference squeezing for the phase
quadratures. The signs of these correlations are interchanged
compared to those achieved in other systems. This fact may be of
importance in applications involving the opto-mechanical coupling
of radiation pressure\cite{Cohadon}. Apart from the simplicity of
our scheme, it also has the potential advantage of being
integrable into existing fibre-optics communication networks.

\begin{figure}[h]
  \begin{center}
 \epsfxsize=2.7 in
 \epsfbox{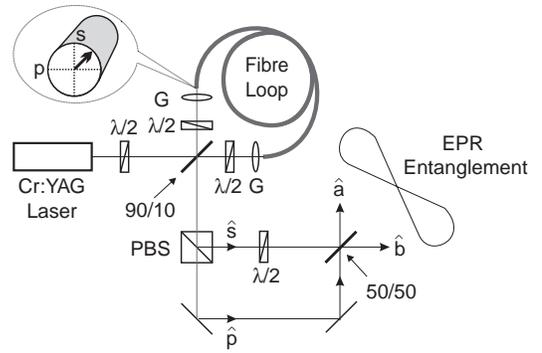}
  \end{center}
  \caption{Schematic of the experimental setup.  $\lambda/2$:
  half-wave plate, G: grin lens, PBS: polarizing beam splitter, 50/50:
  beam splitter with 50\% reflectivity and 90/10: beam splitter with
  90\% reflectivity.  $\hat{s}$ and $\hat{p}$ are the two squeezed
  beams from the respective polarization states.  $\hat{a}$ and
  $\hat{b}$ are the EPR entangled output beams.  Insert shows the
  polarization direction of the input beam to the fibre.}
\label{expt}
\end{figure}

Our experimental setup is as shown in Figure \ref{expt}.  A
passively mode-locked ${\rm Cr^{4+}:YAG}$ laser is used to
produce optical pulses at a centre wavelength of 1505~nm with a
repetition frequency of 163~MHz.  The maximum average power of
the laser is around 95~mW and the pulses have a bandwidth limited
hyberbolic-secant shape with a FWHM of 130~fs.  These pulses are
injected into an asymmetric fibre Sagnac interferometer.  The
Sagnac loop consists of an 8~m long polarization maintaining
fibre (FS-PM-7811 from 3M) and a beam splitter.  The fibre has a
birefringence, characterized by a beatlength of $1.95$~mm for
1505~nm light, that supports the s- and p- polarization states
with negligible cross-talk. The beam splitter of the
interferometer has 91\% (90\%) reflectivity and 9\% (10\%)
transmittivity for the s- (p-) polarization states,
respectively.  This provides a strong and a weak
counterpropagating pulse within the Sagnac loop for each
polarization.

Because of the Kerr nonlinearity of the optical fibre, the strong
pulses acquire an intensity dependent phase shift during propagation.
This mechanism squeezes the circular shaped phase space uncertainty
into an ellipse \cite{Drummond}.  However, the squeezed quadrature of
the ellipse is not aligned with the amplitude quadrature and amplitude
squeezing is not readily observed.  In the asymmetric Sagnac
interferometric arrangement, the weak and the strong
counterpropagating pulses acquire different nonlinear phase shifts.
When the pulses interfere at the beam splitter, this relative phase
shift realigns the axes of the ellipse.  For certain input energies,
this realignment yields direct detectable amplitude squeezing
\cite{Schmitt}.

In contrast to the setup reported by Schmitt {\it et al.}
\cite{Schmitt}, however, we utilized the polarization maintaining
characteristics of the fibre to simultaneously sustain two
orthogonally polarized modes in the Sagnac loop by choosing an input
polarization direction at around 45$^{\circ}$ relative to the fibre
axes.  Thus, two independently amplitude squeezed beams, labelled as
$\hat{s}$ and $\hat{p}$ in Figure \ref{expt}, are produced.

Owing to the birefrigence of the optical fibre, the $\hat{s}$ and
$\hat{p}$ polarization modes of the same energy in general do not
experience the same effective Kerr nonlinearity.  This is because both
the propagation time and the mode confinement for each polarization
are different.  Furthermore, the reflectivity of the Sagnac mirror is
slightly different for the two polarizations.  As a result,
simultaneous squeezing of both modes is not always guaranteed.  This
imbalance can be compensated by careful adjustment of the energy
splitting ratio between the two polarizations.

The photodetectors used are a pair of balanced windowless InGaAs
detectors from Epitaxx (ETX-500).  The detection efficiencies are
measured to be around $(92 \pm 5)$\% and the detection frequency is at
10~MHz with a resolution bandwidth of 300~kHz.  Our detectors have
their 3~dB roll-off frequencies at around 20~MHz and a more than 40 dB
attenuation at the repetition frequency and the harmonics of the
mode-locked ${\rm Cr^{4+}:YAG}$ laser.  In order to obtain an EPR
entanglement, the polarization of the $\hat{s}$ beam and its path
length are adjusted to interfere with the $\hat{p}$ beam at a 50/50
beam splitter.

For two squeezed sources with identical squeezed quadrature and
optical power, EPR entanglement is maximized when the interference
phase is such that the two output beams have equal optical power
\cite{Leuchs}.  We can model this condition by assuming that both
reflected beams have $90^{\circ}$ phase shifts relative to their
transmitted beams.  In order to evaluate the noise variance
expressions of the output beams, we express the annihilation operator
of a field as $\hat{A}(t) = \alpha + \delta \hat{A}(t)$, with $\alpha$
being the classical steady state value, and $\delta \hat{A}(t)$ the
zero-mean operator which contains all the classical and quantum
fluctuations.  We denote the amplitude and phase quadratures with
$\hat{X}^{+}(t) = \hat{A}^{\dagger}(t)+\hat{A}(t)$ and $\hat{X}^{-}(t)
= i (\hat{A}^{\dagger}(t)-\hat{A}(t))$, respectively.  Finally, via a
Fourier transform, the spectral variances $ V(\hat{X}_{A}^{\pm}) =
V(\hat{X}_{A}^{\pm}) (\omega) = \langle |\delta \hat{A}(\omega) \pm
\delta \hat{A}^{\dagger}(\omega)|^{2}\rangle$ are calculated by only
considering the first order contribution of the fluctuation terms.
The noise variances of the output beams $\hat{a}$ and $\hat{b}$ are
calculated to be
\begin{eqnarray}
   V(\hat{X}_{a}^{+}) & = & \frac{1}{4} [V(\hat{X}_{s}^{+}) +
   V(\hat{X}_{s}^{-}) + V(\hat{X}_{p}^{+}) + V(\hat{X}_{p}^{-})]
   \label{abeam} \\
   V(\hat{X}_{b}^{+}) & = & \frac{1}{4} [V(\hat{X}_{s}^{+}) +
   V(\hat{X}_{s}^{-}) + V(\hat{X}_{p}^{+}) + V(\hat{X}_{p}^{-})]
   \label{bbeam},
\end{eqnarray}
and the normalized sum and difference photo currents for both
amplitude and phase quadratures are then given by
\begin{eqnarray}
   V_{\rm sum}^{\pm} = \frac{V(\hat{X}_{a}^{\pm} +
   \hat{X}_{b}^{\pm})}{V(\hat{X}_{a,{\rm SN}}^{\pm} + \hat{X}_{b,{\rm
   SN}}^{\pm})} & = & \frac{1}{2} [V(\hat{X}_{s}^{\pm}) +
   V(\hat{X}_{p}^{\pm})] \label{+beam} \\
   V_{\rm diff}^{\pm} = \frac{V(\hat{X}_{a}^{\pm} -
   \hat{X}_{b}^{\pm})}{V(\hat{X}_{a,{\rm SN}}^{\pm} - \hat{X}_{b,{\rm
   SN}}^{\pm})} & = & \frac{1}{2} [V(\hat{X}_{s}^{\mp}) +
   V(\hat{X}_{p}^{\mp})] \label{+pbeam},
\end{eqnarray}
with the field modes denoted by the respective subscripts, and ${\rm
SN}$ is used to denote the corresponding parameter for a shot noise
limited beam.  For amplitude squeezed input beams, we have
$V(\hat{X}_{s,p}^{+}) < 1 < V(\hat{X}_{s,p}^{-})$, thus giving sum
squeezing for the amplitude quadrature, $V_{\rm sum}^{+} <1$, and
difference squeezing for the phase, $V_{\rm diff}^{-} < 1$.  Apart
from these, we note that all other variances for the output beams are
above the quantum noise limit.

\begin{figure}
  \begin{center}
  \epsfxsize= 2.7 in
 \epsfbox{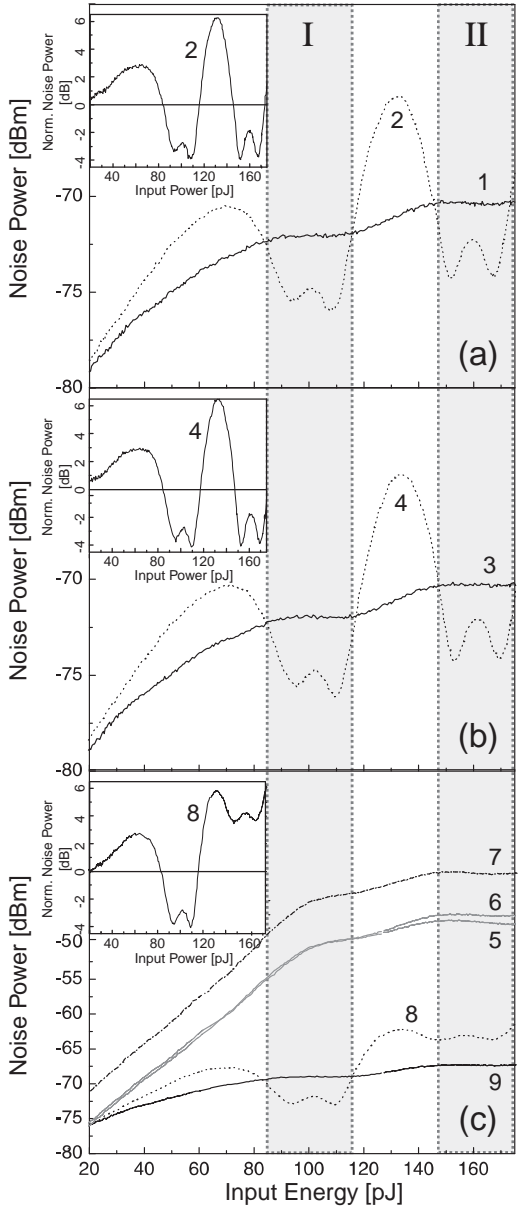}
  \end{center}
\caption{Experimental results.  (a) + (b): squeezing as a
function of input pulse energy for the s-  (p-) polarized beams,
respectively.  (c): noise measurements when both squeezed beams
are made to interfere at the 50/50 beam splitter. Inserts show
the normalized variances of the respective traces. Region I gives
the energy range with EPR entangled outputs; in region II both
input beams appear squeezed, but no quantum correlations are
measured at the outputs (see text).} \label{sqz}
\end{figure}

Our initial experimental investigation is done by varying the
input pulse energy continuously while measuring the $\hat{s}$ and
$\hat{p}$ output beams individually using a balanced detection
scheme.  The results of these measurements are shown in Figure
\ref{sqz} (a) and (b).  Traces 1 and 3 are the respective quantum
noise limits and Traces 2 and 4, the amplitude noise variances.
We observed that similar to \cite{Schmitt}, the amplitude noise
power varies as a function of the input pulse energy giving rise
to squeezing regions I and II. This is a consequence of the
variation in the nonlinear phase shift experienced as a function
of pulse energy.  The noise power traces have a double-dip
structure within a squeezing region due to the non-optimal
reflectivity of the Sagnac beam splitter \cite{Schmitt}.  We note
that the squeezing regions of these two different polarization
beams overlap almost completely.  The maximum amplitude squeezing
obtained in region I is $4.1 \pm 0.2$~dB ($3.9 \pm 0.2$~dB) for
the p- (s-) polarization beam.

In our experiment the beam splitter transforms the two squeezed
input beams into quantum entanglement between the output beams.
The efficiency of this process relies on high interferometric
visibility.  For the input energies of region I a visibility of
$96 \pm 1$\% was measured from a beam splitter with $51.5 \pm
1.0$\% reflectivity.  After the interference the outputs are
measured individually.  In Figure \ref{sqz} (c), Traces 5 and 6
show the amplitude noise variances of the $\hat{a}$ and $\hat{b}$
EPR beams, respectively. As predicted by Eqs.~(\ref{abeam},
\ref{bbeam}), the noise variances of the indi-
\noindent vidual  beams are far above the quantum noise limit
since each of them contains the squeezed and the anti-squeezed
noise variances of both the $\hat{s}$ and $\hat{p}$ beams.  Trace
7 shows that the noise power obtained when both beams are detected
and subtracted is 6~dB above that of the individual EPR beams
(Traces 5 \& 6). This is because it is measured at twice the
optical power of an individual EPR beams (3~dB) and $V_{\rm
diff}^{+} \approx 2 V(X_{a,b}^{+})$ (a further 3~dB).

The excess noise is present in the variances of each individual
beam (Traces 5 \& 6), but the variance of the sum photo currents
(Trace 8) is observed to be below the quantum noise limit (Trace
9). The best result obtained in our setup is found at the input
energy near 110~pJ with $4.0 \pm 0.2$~dB of quantum correlation.
Although there is squeezing of both input beams in region II, the
sum variance of the amplitudes is above the quantum noise limit.
This is attributed to the following reasons.  Firstly there is a
lack of cancellation from the balanced detector pair due to
nonlinear responses at input energy beyond 130~pJ. Moreover, the
interferometric visibility is only optimized for region I at the
input energy of around 110~pJ. At higher input energies,
stimulated Raman scattering begins to dominate, which leads to a
decrease in the spectral overlap between the beams.  Hence, EPR
entanglement is present only in region I (see Fig.  \ref{sqz}).

\begin{figure}[h]
  \begin{center}
\epsfxsize=\columnwidth
 \epsfbox{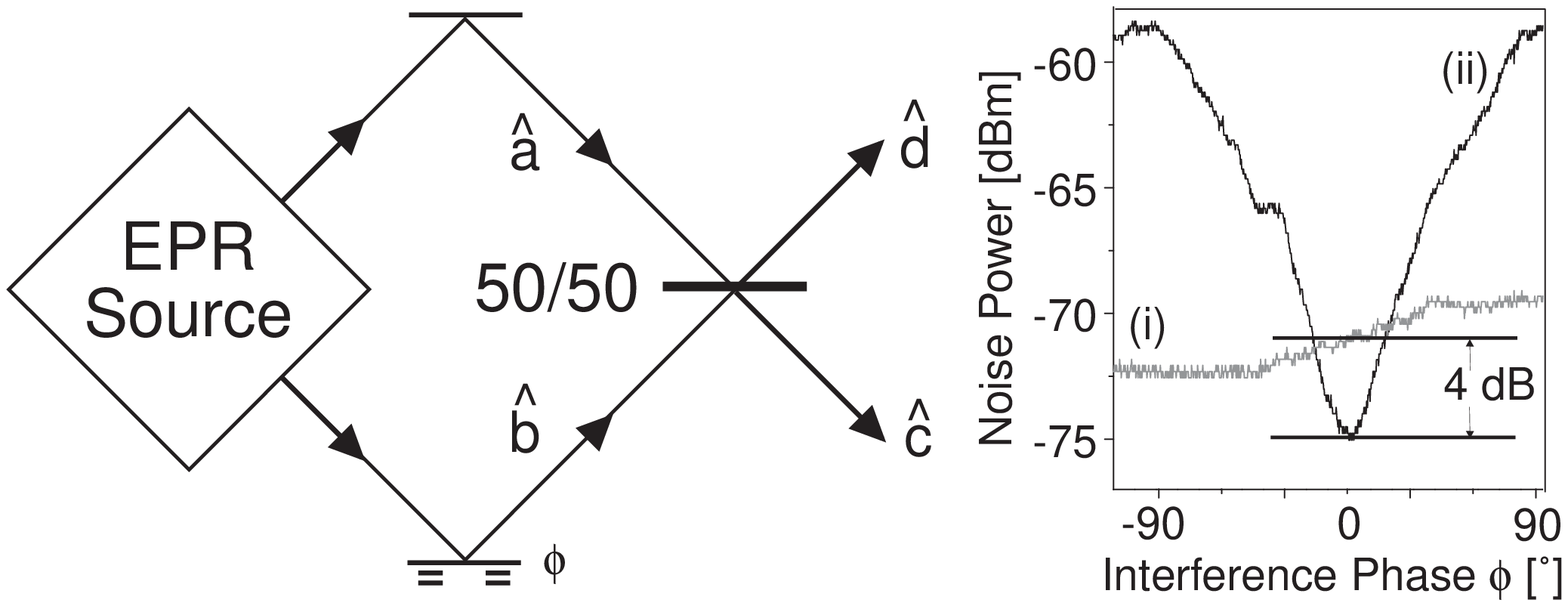}
  \end{center}
\caption{Left: Schematic for the indirect interrogation of the phase
quadrature correlation.  Right: Noise variance of the output beam
$\hat{c}$ when the interferometer phase is scanned.  (i): Quantum
noise limit, (ii): Noise variance of $\hat{c}$.}
\label{phase}
\end{figure}

In order to directly measure the phase quadrature correlation of the
output beams, optical local oscillators are required for homodyne
measurements.  In our experiment, this is not possible due to pump
power limitation and detector saturation.  Instead, an indirect
interrogation of the phase quadrature correlation as shown in Figure
\ref{phase} is devised.  Let us assume that the experiment described
thus far is a black box with two output beams of which the amplitude
correlation has been established.  We now let these beams interfere at
yet another 50/50 beam splitter.  The final output beams are now
denoted by $\hat{c}$ and $\hat{d}$ and their variances are given by
\begin{eqnarray}
    V(\hat{X}_{c,d}^{+}) & = & \frac{1}{4} \langle ( \delta \!
    \hat{X}_{a}^{+} + \delta \!  \hat{X}_{b}^{+} + \delta \!
    \hat{X}_{a}^{-} - \delta \!  \hat{X}_{b}^{-} )^{2} \rangle \\
    & = & \frac{1}{4} \left [V\!  (\hat{X}_{a}^{+} + \hat{X}_{b}^{+} )
    + V\!  (\hat{X}_{a}^{-} - \hat{X}_{b}^{-} ) \nonumber \right ] \\
    & & \phantom{ \alpha^{2}} \pm \left (\langle \delta \!
    \hat{X}_{a}^{+} \ \delta \!  \hat{X}_{a}^{-} \rangle - \langle
    \delta \!  \hat{X}_{b}^{+} \ \delta \!  \hat{X}_{b}^{-} \rangle
    \right ) \nonumber \\
   & & \phantom{ \alpha^{2}} \pm \left (\langle \delta \!
   \hat{X}_{b}^{+} \ \delta \!  \hat{X}_{a}^{-} \rangle- \langle
   \delta \!  \hat{X}_{a}^{+} \ \delta \!  \hat{X}_{b}^{-}\rangle
   \right ) \label{inferphase}
\end{eqnarray}
By argument of symmetry, we note that $\hat{a}$ and $\hat{b}$ are
produced similarly and interchanging the indices should not give a
different result.  The terms within each bracket of the second and
third lines of Eq.~(\ref{inferphase}) therefore should cancel with
each other.  Hence measurements of the amplitude variances of
$\hat{c}$ and $\hat{d}$ are equivalent to combined measurements
of the amplitude sum variance, $V (\hat{X}_{a}^{+} +
\hat{X}_{b}^{+}) $, and the phase difference variance, $V
(\hat{X}_{a}^{-} - \hat{X}_{b}^{-}) $.  Since $V (\hat{X}_{a}^{+}
+ \hat{X}_{b}^{+}) $ is known from earlier measurements, we can
therefore deduce $V (\hat{X}_{a}^{-} - \hat{X}_{b}^{-}) $ without
using a separate local oscillator.  With this method we measured
a best difference squeezing of $V_{\rm diff}^{-} = 4.0 \pm
0.4$~dB for the phase quadratures at the input pulse energy of
110~pJ, in excellent agreement with the prediction of
Eq.~(\ref{+pbeam}).  The result of a typical scan of this second
interferometer phase is shown in Fig.~\ref{phase}.  The noise
suppression predicted by Eq.~(\ref{inferphase}) is only observed
at the interferometer phase of $\phi = 0$, corresponding to
$\hat{c}$ and $\hat{d}$ having equal power.

In the experiment the quantum correlations in the two
non-commuting observables, amplitude and phase are

\noindent  characterized by the sum $ V_{\rm sum}^{+} = 0.40 \pm
0.02 $ and difference $ V_{\rm diff}^{-} = 0.40 \pm 0.04 $
squeezing of the entangled beams where the corresponding
variances of the sum and difference photo currents are normalized
to the shot noise level of both beams
(Eqs.~\ref{+beam},\ref{+pbeam}). With this normalization the
Peres-Horodecki separability criterion for continuous variables
\cite{characterization} can be expressed as $\label{separability}
V_{\rm sum}^{+} + V_{\rm diff}^{-} < 2$. Thus the value of $
V_{\rm sum}^{+} + V_{\rm diff}^{-} = 0.80 \pm 0.03 < 2$ shows the
high nonseparability of the generated state. For many quantum
information applications, this criterion also defines the
boundary between the classical and the quantum regime. For
example, the limit of the Peres-Horodecki criterion corresponds
to a teleportation fidelity of $ F = \langle \psi_{\rm in}
|\rho_{out} | \psi_{\rm in} \rangle = 0.5$, which separates
classical and quantum teleportation. Assuming our realistic
EPR-source and ideal conditions for the teleportation, this would
yield a fidelity of $F = 0.71\pm 0.02 > \frac{2}{3}$ (for
criteria of quantum teleportation see \cite{Braunstein},
\cite{Ralph}, \cite{Grangier}).

In terms of demonstrating the continuous variable EPR-{\it
Gedankenexperiment}, one has to apply another criterion.  The
demonstration of the EPR-paradox requires the ability to infer
``at a distance" both non-commuting observables with a precision
below the quantum noise limit of {\it one single beam}
\cite{Reid}. The symmetry between the beams in optical powers and
quantum uncertainties allows us to renormalize to obtain the
appropriate inferred variances for the amplitude and phase
quadrature. These inferred variances for an optimum gain of unity
are then the conditional variances . Their product
$V(\hat{X}^{+}_{a, {\rm cond}}) V(\hat{X}^{-}_{a, {\rm cond}}) =
V (\hat{X}_{a}^{+} + \hat{X}_{b}^{+}) V (\hat{X}_{a}^{-} -
\hat{X}_{b}^{-}) = 0.64\pm 0.08$ is less than unity satisfying
the condition for the demonstration of the EPR paradox
\cite{Ou,Reid}.

In summary, we have presented a scheme for the generation of EPR
entangled beams using the Kerr nonlinearity of an optical fibre.
Apart from subtracting the electronic noise of the detectors, all
results presented are observed quantities not corrected for
linear losses such as detection efficiency.  Moreover, the
stability of the EPR entanglement is limited only by the
stability of the pump laser. Since the fibre is the only
nonlinear medium used, it has advantages over the existing
methods of EPR entanglement generation in terms of its
simplicity, cost and integrability into existing communication
networks. There are two possible outlooks for this basic building
block for quantum communication.  Firstly, there is some prospect
to further improve the quality of the entanglement, since more
than 11~dB of amplitude squeezing was predicted even when
realistic experimental parameters were considered with a 93/7
Sagnac beam splitter \cite{Schmitt}. Secondly, the lack of the
phase measurement can be overcome by implementing an arm length
unbalanced interferometer, which under appropriate conditions
converts   the phase quadrature variances into directly
detectable amplitude variances without introducing additional
noise. Then, due to its simplicity, multiple EPR entanglements
can be built and other more sophisticated quantum communication
schemes can be demonstrated in the future.

We gratefully acknowledge financial support of the Deutsche
Forschungsgemeinschaft and of the EU grant under QIPC, project
IST-1999-13071 (QUICOV).  P.~K.~Lam is an Alexander von Humboldt
fellow.


\end{document}